\documentclass[preprint,preprintnumbers,amsmath,amssymb,floatfix,nofootinbib,h-physrev.bst]{revtex4}
\usepackage{epsf, array, color}
\usepackage{graphicx}
\usepackage{subfigure}
\usepackage{bbold}
\usepackage{subfig}
\usepackage{amsmath}
\usepackage{mathtools}
\usepackage{dcolumn}
\usepackage{url}
\usepackage{verbatim}
\usepackage{wrapfig}
\usepackage{amsfonts}
\usepackage{comment}
\usepackage{epigraph}
\usepackage{slashed}
\usepackage[utf8]{inputenc}
\usepackage{bbm}
\usepackage{cancel}
\usepackage{epsf, array, color}

\newcommand{\bea}{\begin{eqnarray}}
\newcommand{\eea}{\end{eqnarray}}

\def\beq{\begin{equation}}
\def\eeq{\end{equation}}

\newcommand{\lsim}{\lesssim}

\begin{document}

\title{Enhanced di-Higgs Production in the Complex Higgs Singlet Model }

\author{S.~Dawson$^{\, a}$ and M.~ Sullivan$^{\, a,b}$ }
\affiliation{
\vspace*{.5cm}
  \mbox{$^a$Department of Physics,\\
  Brookhaven National Laboratory, Upton, N.Y., 11973,  U.S.A.}\\
 \mbox{$^b$ Department of Physics and Astronomy, University of Kansas, Lawrence, Kansas, 66045 USA}
 \vspace*{1cm}}

\date{\today}

\begin{abstract}
We consider the Standard Model (SM) extended by  the addition of a complex scalar singlet, with no assumptions about 
additional symmetries of the 
potential.  This model provides for resonant di-Higgs production of Higgs particles with different masses. 
 We demonstrate
that regions of parameter space allowed by precision electroweak measurements, experimental limits on single Higgs production, 
and perturbative unitarity allow for large di-Higgs production rates relative to the SM rates.  In this scenario, 
 the dominant production mechanism  of the new scalar states is di-Higgs production.  Results are presented for
$\sqrt{S}=13$, $27$ and $100~TeV$.
\end{abstract}

\maketitle

\section{Introduction}

The exploration of the Higgs sector is a primary focus of the LHC physics program, with measurements of the Higgs couplings
to fermions and gauge bosons, the Higgs mass, and Higgs CP properties becoming ever more precise.  Very little is known, however, about the 
Higgs tri-linear and quartic self-couplings which are unambiguously predicted in the Standard Model (SM).  The  SM
Higgs tri-linear coupling can be most sensitively probed by double Higgs production 
through gluon fusion which unfortunately has a very 
small rate\cite{Plehn:1996wb}, even at high energy and high luminosity\cite{Frederix:2014hta}.  The
best current limit  on double Higgs production is from the ATLAS experiment\cite{ATLAS-CONF-2016-049},
 $\sigma(pp\rightarrow hh)/\sigma(pp\rightarrow hh)_{SM} < 29$, with prospects for  only modest
improvements at higher luminosity.   A definitive measurement of the SM  tri-linear Higgs self-coupling appears out of 
reach at the LHC\cite{Baur:2002qd,Dolan:2012rv,Baglio:2012np}. 

Given the small SM rate for double Higgs production, it is an excellent place to search for Beyond the SM  (BSM) physics.  In the presence of a 
scalar resonance coupling to the  SM-like Higgs boson, the double Higgs rate can be
significantly enhanced.  
This can occur in the MSSM and the NMSSM, for example. 
The simplest  possibility is to add a hypercharge-$0$ real scalar to the model 
which interacts with SM fermions and gauge bosons only through the mixing with the Higgs doublet.
The LHC phenomenology in the context of the real singlet model has been extensively studied in the literature\cite{Barger:2014taa,Barger:2007im,Profumo:2007wc,Pruna:2013bma,Chen:2014ask,Dawson:2015haa,Lewis:2017dme,No:2013wsa}.  When the most general 
scalar potential (without the imposition of a $Z_2$ symmetry) is considered,
the real singlet model can have a first order electroweak phase transition\cite{Espinosa:2011ax,Profumo:2014opa,Curtin:2014jma,Chen:2017qcz} for some
values of the parameters.  

The complex scalar singlet extension has new features beyond the real singlet case.  It has several phases, 2 of which can accommodate a dark matter 
candidate\cite{Coimbra:2013qq,Gonderinger:2012rd}. In  the
broken phase of this model (which is the subject of this work) there are $3$ neutral scalar particles which  mix to form the mass eigenstates, one of which is the $125~GeV$ scalar.   Final states with $2$ different mass scalar
 particles can be resonantly produced in this scenario and there are large regions of parameter space where the couplings of the new
 scalars to SM particles are highly suppressed, making the dominant production mechanism 
 of the new scalars the 
 Higgs decays to other Higgs-like particles.  The resonant
 production of two different mass Higgs particles is a smoking gun for this class of theories.    

We study the most general case of a complex scalar singlet extension of the SM, without the introduction of any new symmetries
for the potential.  The
complex singlet model has
been previously studied imposing a softly broken $U(1)$ symmetry and benchmark points described for the study of the decay of the
heavy scalar to the SM Higgs boson and the lighter scalar of the model\cite{Costa:2015llh,Muhlleitner:2017dkd}. 
The parameter space of the model we study is larger, allowing for new 
phenomenology.  The basic features of the model are discussed in Section II and 
  the limits  on the  model from perturbativity, unitarity and the oblique parameters are presented  in Sec.
\ref{sec:lims}.  Our most interesting results are  the implications for double Higgs studies  and the description
of scenarios where one of the new Higgs bosons is predominantly produced in association with the $125~GeV$ boson.
This is  discussed in Sec. \ref{sec:hh}.

\section{Model}
\label{sec:model}
We consider a model containing the SM $SU(2)$ doublet, $\Phi$, and a complex scalar singlet, $S_c$.
Since $S_c$ has hypercharge -$0$ it does not couple directly to SM fermion or gauge fields, and its  tree level interactions 
with SM fermions and gauge bosons result entirely from mixing with $\Phi$.
The most general renormalizable scalar potential is\cite{Barger:2008jx},
\begin{eqnarray}
{\cal V}(\Phi,S_c)&=&
{\mu^2\over 2}\Phi^\dagger\Phi+{\lambda\over 4}(\Phi^\dagger\Phi)^2
+\biggl({1\over 4} {\delta_1}
\Phi^\dagger\Phi S_c
+{1\over 4}\delta_3 \Phi^\dagger\Phi S_c^2
+a_1S_c
\nonumber \\ &&
+{1\over 4} b_1 S_c^2
+{1\over 6} e_1S_c^3
+{1\over 6} e_2S_c\mid S_c\mid^2
+{1\over 8}d_{1}S_c^4
+{1\over 8}d_{3}S_c^2\mid S_c\mid^2
+ h.c.\biggr)
\nonumber \\ &&
+{1\over 4}d_2(\mid S_c\mid^2)^2
 +{\delta_2\over 2} \Phi^\dagger \Phi \mid S_c\mid^2
 +{1\over 2}b_2\mid S_c\mid^2
 \, ,
\label{eq:vdef}
\end{eqnarray}
where $a_1,b_1,e_1,e_2,d_1,d_3, \delta_1$ and $\delta_3$ are complex. 
After spontaneous symmetry breaking, in unitary gauge, 
\begin{equation}
\Phi=\left(\begin{matrix}0\\{h+v\over\sqrt{2}}\end{matrix}\right),\quad S_c={1\over\sqrt{2}}\biggl(S+v_S+i(A+v_A)\biggr)\, .
\end{equation}
Since we have included all allowed terms in Eq. \ref{eq:vdef}, the coefficients can always be redefined such that
$v_S=v_A=0$. This makes the potential of Eq. \ref{eq:vdef}
identical to that obtained by adding $2$ real singlets to the SM and there is no CP violation.   Previous work\cite{Costa:2015llh,Barger:2008jx}
 imposed a global $U(1)$ symmetry or a $Z_2$
symmetry to eliminate some of the terms in the potential, making the shift to $v_S=v_A=0$  in general not possible. 

The mass eigenstate fields are $h_1,h_2,h_3$  (masses $m_1,m_2,m_3$) are found from the rotation,
\begin{equation}
\left( 
\begin{matrix}
h_1\\
h_2\\
h_3\\
\end{matrix} 
\right)= {V}
\left( \begin{matrix}
h\\
S\\
A\\
\end{matrix}
\right)\, ,
\label{eq:massbasis}
\end{equation}
where $V$ is a $3\times 3$ unitary matrix with,
\begin{equation}
{V}\equiv\left (
\begin{matrix}
c_1 & -s_1 c_3 & -s_1s_3\\
s_1c_2 & c_1c_2 c_3- s_2 s_3 & c_1c_2s_3+s_2c_3 \\
s_1s_2 & c_1s_2c_3+c_2s_3 & c_1s_2s_3-c_2c_3 \\
\end{matrix}\right)
\label{eq:ckm}
\end{equation}
and we abbreviate $c_i=\cos\theta_i$, etc. 
Note that the phase usually associated with the CKM-like mixing matrix does not appear since the mass matrix in terms of
the  real fields $h$, $S$, and $A$ is strictly real by hermiticity.  Since all allowed terms are included in Eq. \ref{eq:vdef}, we are free to perform a field redefinition $S_c \rightarrow S_c e^{i\phi}$ while leaving the form of the potential unchanged. We choose to take $S_c \rightarrow S_c e^{i\theta_3}$. This results in the field redefinitions,
\begin{equation}
\label{eq:phaserot}
\left( 
\begin{matrix}
h\\
S\\
A\\
\end{matrix} 
\right)\rightarrow \left( \begin{matrix}
1 & 0 & 0 \\
0 & c_3 & -s_3 \\
0 & s_3 & c_3 \\
\end{matrix} \right)
\left( \begin{matrix}
h\\
S\\
A\\
\end{matrix}
\right)\, ,
\end{equation}
which, when combined with Eqs.~\ref{eq:massbasis} and \ref{eq:ckm} with matrix multiplication, leads  to a simplified mixing matrix,
\begin{equation}
{V} \rightarrow \left (
\begin{matrix}
c_1 & -s_1 & 0\\
s_1c_2 & c_1c_2&s_2 \\
s_1s_2 & c_1s_2 & -c_2 \\
\end{matrix}\right)\, .
\label{eq:ckmreduced}
\end{equation}
So we see that performing a suitable phase rotation is equivalent to setting $\theta_3=0$. For the rest of the paper, we use this convention to eliminate $\theta_3$.

We take as inputs to our scans,
\begin{equation}
v=246~GeV, m_1=125~GeV, m_2,m_3,\theta_1, \theta_2, \delta_2, \delta_3, d_1, d_2, d_3, e_1, e_2\,
\label{eq:parms}
\end{equation}
where $\delta_3,d_1,d_3,e_1$ and $e_2$ can be complex and are defined in Eq. \ref{eq:vdef}.

The SM-like Higgs boson is identified with $h_1$ with $m_1=125~GeV$.  The couplings of $h_1$ to SM particles
are suppressed by a factor $c_1$ relative to the SM rate.  
The states are ordered according to their couplings to SM particles.  $h_1$ has the strongest couplings to SM particles,
$h_2$ couplings are suppressed by $s_1c_2$ relative to the SM couplings, and $h_3$ couplings are the smallest,
and are suppressed by $s_1s_2$ relative to SM couplings.  The mass ordering of $h_2$ and $h_3$ is arbitrary. 
The ATLAS experiment restricts the value of $c_1$ to be,
\begin{equation}
c_1=\mid V_{11}\mid~ >~ 0.94\, ,
\end{equation}
at $95\%$ confidence level using Run-1 Higgs coupling fits\cite{Aad:2015pla}.  Similarly, a global fit to Higgs coupling strengths
 by CMS and ATLAS\cite{Khachatryan:2016vau},
\begin{equation}
\mu=1.09\pm .11\, , 
\end{equation}
yields an identical limit on $c_1$. 

\section{Limits from Perturbativity, Oblique Parameters and Unitarity}
\label{sec:lims}

\begin{figure}
  \centering
{\includegraphics[width=0.48\textwidth]{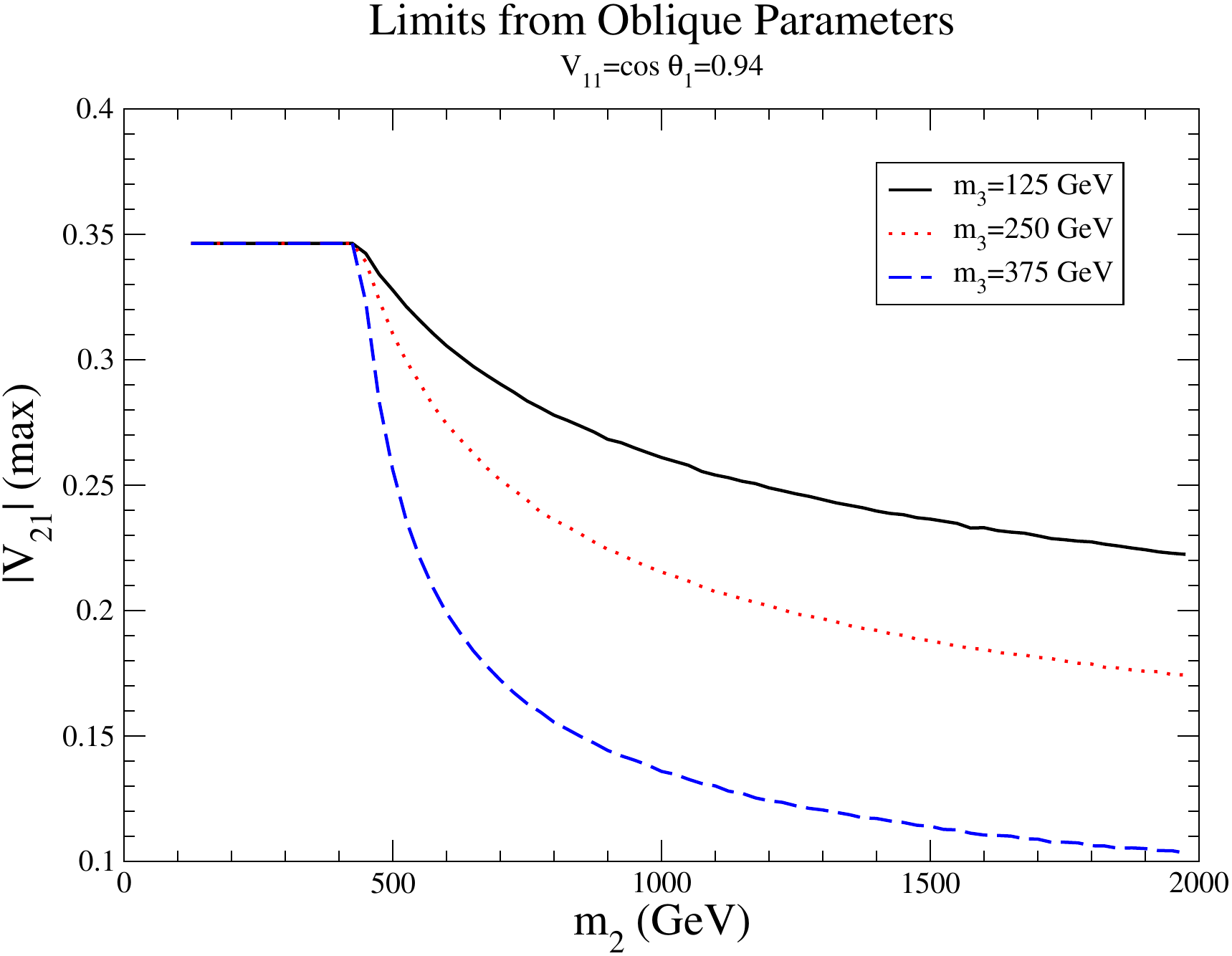}}
{\includegraphics[width=0.48\textwidth]{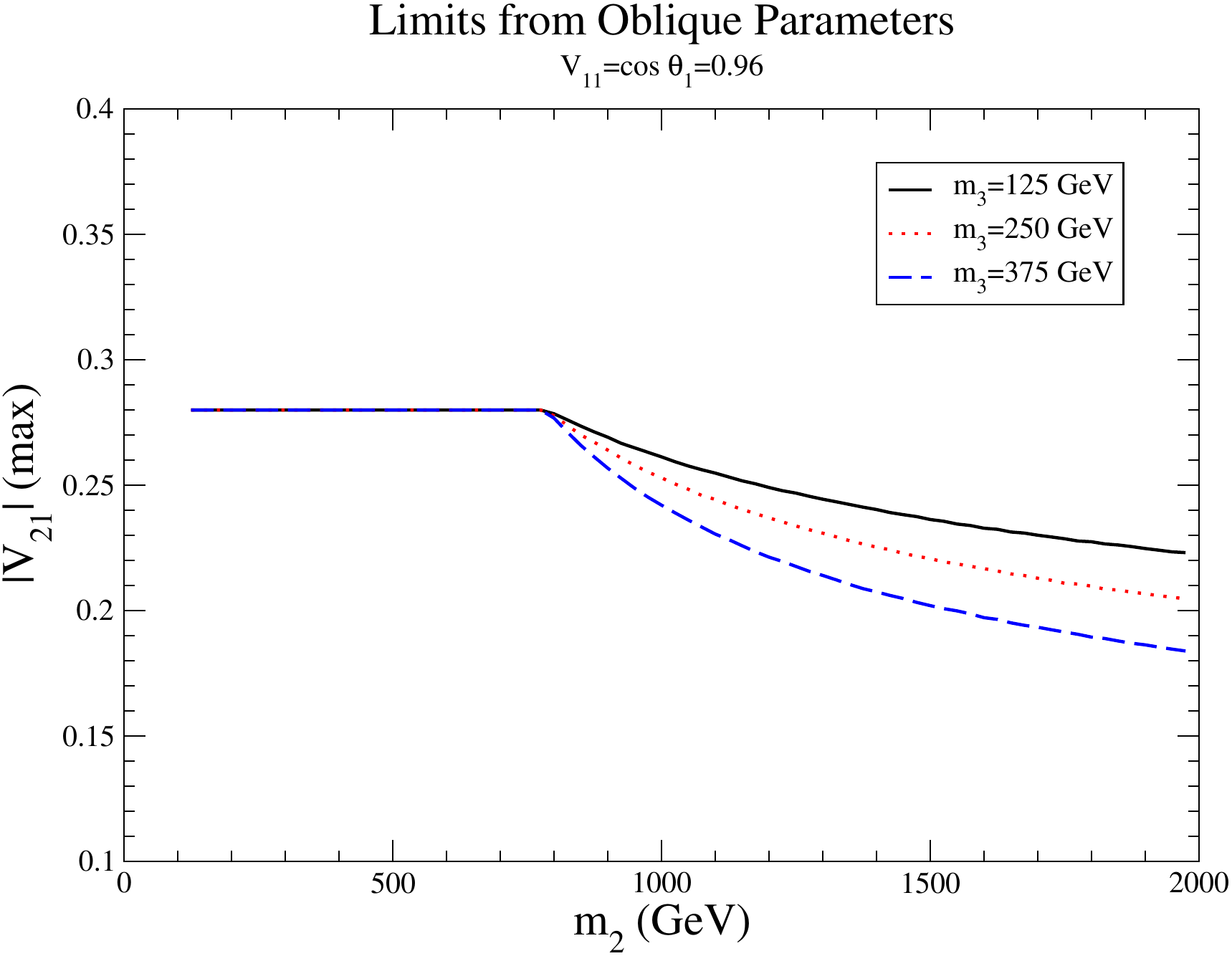}}
 \caption{Limits on $m_2$ for allowed couplings of 
 $h_1$ to SM particles[$\cos\theta_1=.94$ (LHS) and $\cos\theta_1=.96$ (RHS)] for various values of $m_3$ using the oblique parameter 
 ($\cal {S,T,U}$) limits of Ref. \cite{deBlas:2016ojx}.
 \label{fg:stu}
 }
 \end{figure}

The parameters of the model must satisfy constraints from electroweak precision measurements, searches for heavy Higgs bosons,
and limits from perturbative unitarity, along with the restrictions from single Higgs production discussed in the previous section.     
Fits to the oblique parameters place strong limits on the allowed scalar masses and mixings.
 Analytic results for a
model with 2 additional scalar singlets are given in Ref. \cite{Dawson:2009yx}.
For $m_i >> M_W, M_Z$,  the approximate contributions are ,
\begin{eqnarray}
\Delta {\cal S}&\sim&(
1-\mid V_{11}\mid^2) {\cal S}_{SM} +{1\over 12 \pi}
\Sigma_{i=1,2,3}\mid V_{i1}\mid^2
\log\biggl({m_i^2\over m_1^2}\biggr)
\nonumber \\
\Delta {\cal T}&\sim &(1-\mid V_{11}\mid^2){\cal T}_{SM}  -{3\over 16 \pi c_W^2}
\Sigma_{i=1,2,3}\mid V_{i1}\mid^2\log\biggl({m_i^2\over m_1^2}\biggr)
\nonumber \\
\Delta {\cal U}&\sim &(1-\mid V_{11}\mid^2){\cal U}_{SM} 
\, .
\label{eq:loglim}
\end{eqnarray}
The restrictions  from the oblique parameters\cite{deBlas:2016ojx} on 
$V_{21}=s_1c_2$ for the minimum value of $c_1=.94$ allowed by single Higgs production are shown 
on the LHS and for $c_1=.96$ on the RHS of Fig. \ref{fg:stu}.  
TeV scale masses require
quite small values of $V_{21}$, which is the parameter that  determines the coupling of $h_2$ to SM particles.   
 The flat portions of the curves for small $m_2$ in Fig. \ref{fg:stu} represent the imposed limit on $\theta_1$ from
 single Higgs production.  As this limit becomes stronger, the limits from oblique parameters becomes less and less
 relevant. 
 As the $h_1$ couplings become more
and more SM-like ($\theta_1\rightarrow 0$), the allowed coupling of $h_2$ to SM particles becomes highly suppressed.
The constraints from the oblique parameters shown in Fig. \ref{fg:stu} are consistent with those obtained
in the real singlet model in Ref. \cite{Pruna:2013bma}.  For the  values of $\theta_2$ allowed by  Fig. \ref{fg:stu}, the direct searches, $pp\rightarrow
h_{2}(h_3) \rightarrow W^+W^-$ do not provide additional restrictions on $V_{21}$\cite{Aaboud:2017gsl,Khachatryan:2015cwa}.

 \begin{figure}
  \centering
{\includegraphics[width=0.48\textwidth]{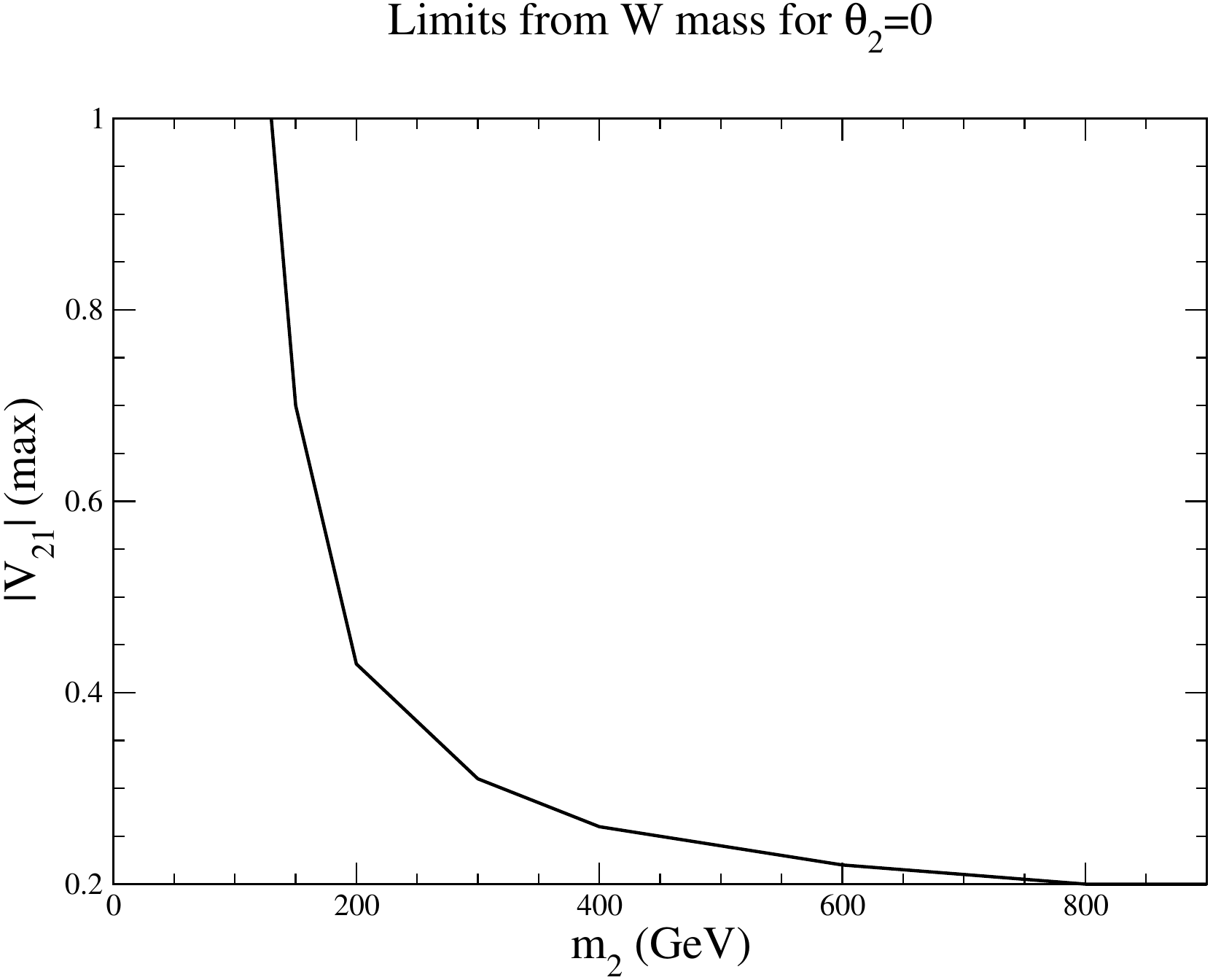}}
 \caption{Maximum allowed value of $V_{21}$  from the $W$ mass measurement as a function of $m_2$ 
 in the real singlet model and in the complex singlet model with $\theta_2=0$
 from Ref. \cite{Lopez-Val:2014jva}. \label{fg:wmass}
 }
 \end{figure}

 In the real singlet model, much stronger constraints are placed on the parameters from the $W$ boson mass than from the 
 oblique parameters\cite{Chalons:2016lyk,Lopez-Val:2014jva}.  For example,  in the real singlet model for $m_{2}=1~TeV$, the $W$
 mass measurement requires $\mid V_{21}\mid < .19$.  For $\theta_2=0$,
 $h_3$ does not couple to SM particles and the results of Refs. \cite{Chalons:2016lyk,Lopez-Val:2014jva} can be applied directly
 to the complex singlet case.  The results of Ref. \cite{Lopez-Val:2014jva} are shown in Fig. \ref{fg:wmass}.  The calculation of the
 limit from the $W$ mass in the complex singlet model for non-zero $\theta_2$ is beyond the scope of this paper and involves
 contributions from all $3$ Higgs bosons and  could potentially yield 
 interesting limits.  The limits from the oblique parameters in the complex singlet case, (Fig. \ref{fg:stu}), demonstrates that the dependence of
 the limits on $m_3$ is non-trivial.

 The quartic couplings in the potential are strongly limited by the requirement of perturbative unitarity of the
 $2\rightarrow 2$ 
 scattering processes\cite{Lee:1977eg}. 
 We compute the $J=0$ partial waves, $a_0$, in the high energy limit where only the quartic couplings contribute and 
 require $\mid a_0\mid < {1\over 2}$. The contributions from the tri-linear couplings are suppressed at
 high energy and do not contribute in this limit.
 For example, we find the restriction from the process, $(SS)/ \sqrt{2}\rightarrow (SS) / \sqrt{2}$,
 \begin{equation}
 Re(d_1+d_2+d_3)\lsim {32\pi\over 3}\, .
 \end{equation}
 Similarly, from $hS\rightarrow hS$, we find, 
 \begin{equation}
 Re(\delta_2+\delta_3) \lsim16 \pi\, .
 \end{equation}
 Looking at the eigenvectors for neutral CP even scattering processes, 
 \begin{equation}
 \biggl\{ 
 \omega^+\omega^-\,, {zz\over \sqrt{2}}\, ,
 {hh\over \sqrt{2}}\,
 ,hS\, ,
 hA\, ,
 {SS\over \sqrt{2}}\,,
 {AA\over \sqrt{2}}\,, A S\biggr\}
\, ,
 \end{equation}
 ($\omega^\pm,z$ are the Goldstone bosons),
   we find the generic 
 upper limits on the real and imaginary  quartic couplings, 
 \begin{eqnarray}
 Re(d_i), Im(d_i)&\lsim & {32\pi\over 3}\,, i=1,2,3\nonumber \\
 \delta_2, Re(\delta_3), Im(\delta_3) \lsim 16 \pi\, .
 \end{eqnarray}
 These upper limits are conservative bounds, and more stringent bounds are obtained from looking at the eigenvalues of the 8 by 8 scattering matrix. These upper bounds on the parameters involve finding solutions to higher order polynomials and do not have simple analytic solutions. Thus, the bounds from perturbative unitarity are determined numerically and imposed in the scans of the next section. 
 
\begin{figure}
  \centering
{\includegraphics[width=0.48\textwidth]{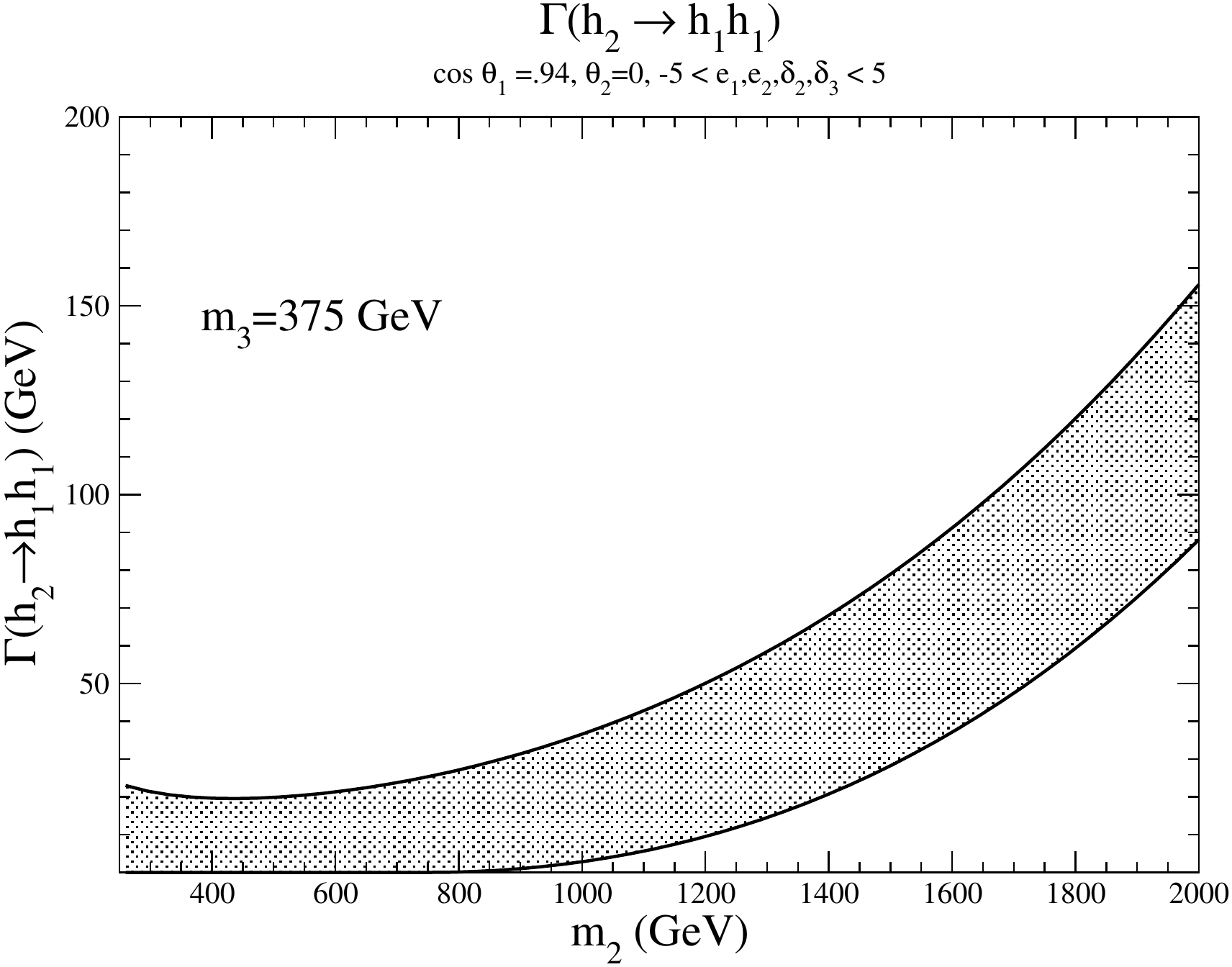}}
 \caption{
  Decay width for $h_2\rightarrow h_1 h_1$ when all parameters
 are taken real and $\delta_2$ and $\delta_3$ are scanned over. \label{fg:lamfig}
 }
 \end{figure}

 The tri-linear Higgs couplings depend on the scalar masses and could potentially become large.  In the limit of small
mixing, $\theta_1<<1$ and $\theta_2=0$, the $h_2 h_1 h_1$ coupling is,
\begin{equation}
\lambda_{211}\rightarrow \sin\theta_1\biggl\{
{2m_1^2\over v}
\biggl(1+{m_2^2\over 2 m_1^2}\biggr)
-v\biggl(\delta_2+Re(\delta_3)\biggr)
\biggr\}\, ,~
\text{small angle limit}
\label{eq:small}
\end{equation}
 and we see that the growth of $\lambda_{211}$ with large $m_2$ is mitigated by the $\sin(\theta_1)$ suppression.   
 The decay width for $h_2\rightarrow h_1h_1$ is\cite{Chen:2014ask},
 \begin{equation}
 \Gamma(h_2\rightarrow h_1 h_1)
 ={\lambda_{211}^2\over 32 \pi m_2}
 \sqrt{1-{4m_1^2\over m_2^2}}\, .
 \end{equation}
 In Fig. \ref{fg:lamfig}, we have  taken all parameters real and scanned over $-5 < \delta_2,\delta_3<5$ for fixed $m_3$, $\theta_1$
 and $\theta_2$.  The dependence on $e_1$ and $e_2$ is minimal in the small angle limit, as demonstrated in 
 Eq. \ref{eq:small}.  In all cases, we have $\Gamma(h_2\rightarrow h_1 h_1) << m_2$, showing that there is no problem
 with the tri-linear couplings becoming non-perturbative in the small angle limit. Increasing the range we scan over changes the numerical
 results, but $\Gamma(h_2\rightarrow h_1 h_1)/m_2$ is always $<<1$. 
  
 Finally, we require that the parameters correspond to an absolute minimum of the potential.  This has been extensively studied for the real singlet
 model in Refs. \cite{Lewis:2017dme,Espinosa:2011ax,Coimbra:2013qq}  and analytic results derived.  For the case of the complex singlet, we scan over parameter space for numerically allowed values of the parameters\cite{Gonderinger:2012rd} and do not obtain an analytic solution. 

\section{Results}
\label{sec:hh}

In the limit of $\theta_2 \rightarrow 0$, (as suggested by the single Higgs rates), the scalar $h_3$ does not couple directly to SM particles and it can only be observed through di-Higgs production. We will consider
$h_3$ to be in the $100-400~GeV$  mass range.
The largest production rate at the LHC is through the resonant process $gg\rightarrow h_2 \rightarrow h_1 h_3$. The complex
singlet model is thus
an example of new physics that will first be seen in the study of di-Higgs resonances\cite{Muhlleitner:2017dkd,Bowen:2007ia}.
We perform a scan over the parameters of Eq. \ref{eq:parms}, subject to the restrictions discussed in the previous 
section\footnote{For the complex singlet model with a $U(1)$ symmetry,
 a comparable scan can be performed using the program ScannerS\cite{Coimbra:2013qq}.}.  We always
fix $c_1=0.94$ and consider the $2$ cases, $\theta_2=0$ and $\theta_2={\pi\over 12}$. 

\begin{figure}
  \centering
{\includegraphics[width=0.48\textwidth]{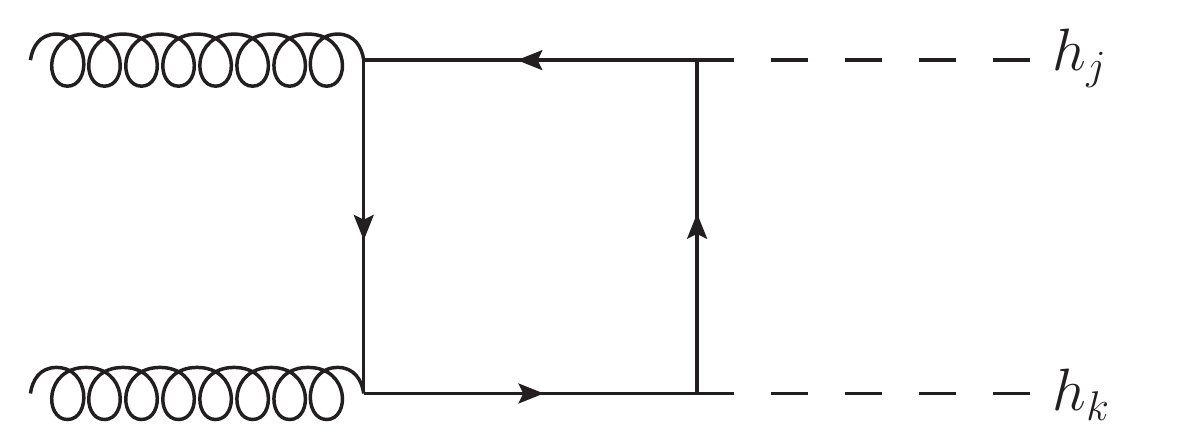}}
{\includegraphics[width=0.48\textwidth]{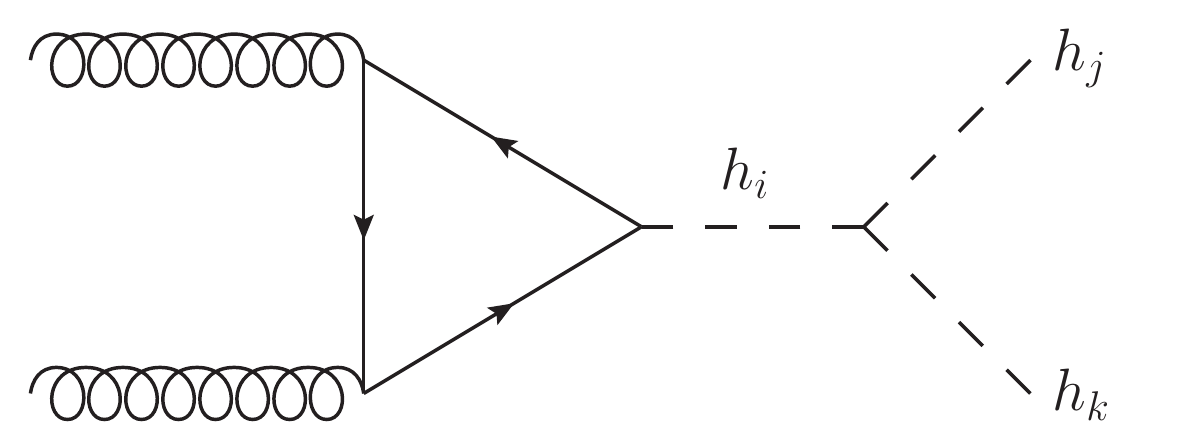}}
 \caption{ Feynman diagrams for the production of $h_j h_k$, $i,j,k=1,2,3$.
 \label{fg:diags}
 }
 \end{figure}

\begin{figure}
  \centering
{\includegraphics[width=0.48\textwidth]{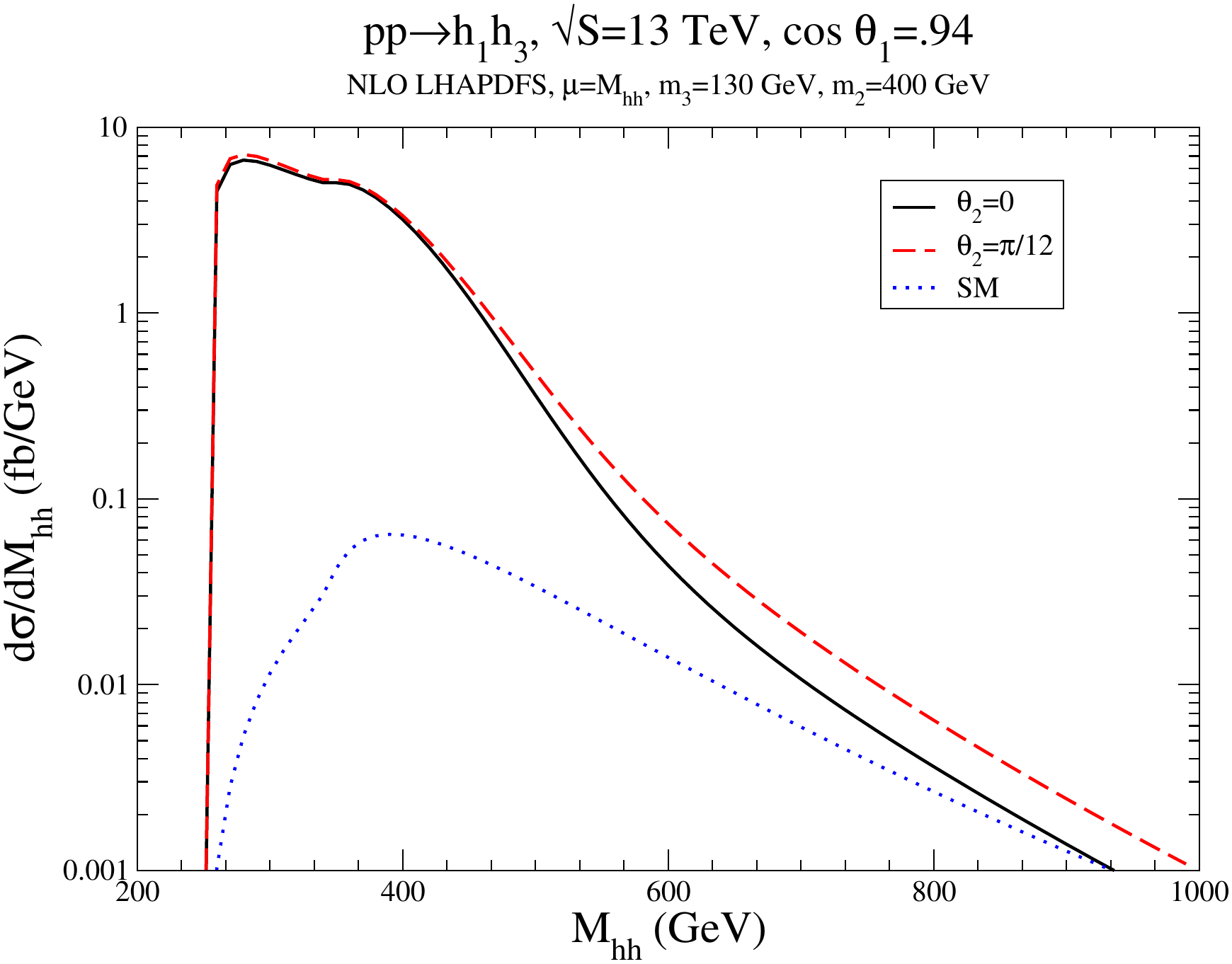}}
{\includegraphics[width=0.48\textwidth]{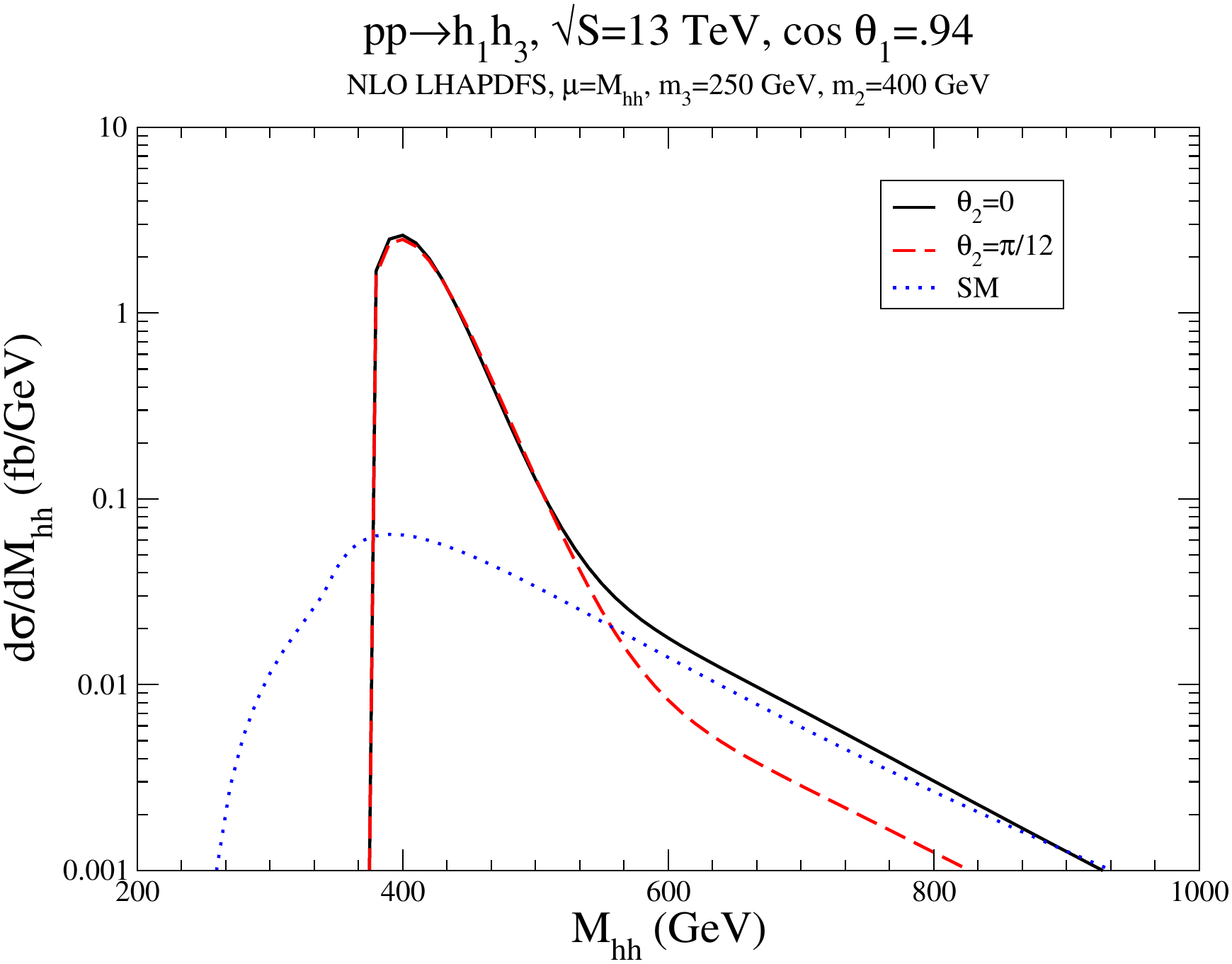}}
 \caption{ $M_{hh}$ spectrum of the complex singlet model production of $h_1h_3$ from the resonant exchange of $h_2$.
 The dominant contribution in the loops is from the top quark. 
 \label{fg:mhh}
 }
 \end{figure}

For the allowed parameter space, we compute the amplitude for $gg\rightarrow h_1 h_3$ shown in Fig. \ref{fg:diags}.  Analytic results
in the context of the MSSM are given in Ref. \cite{Plehn:1996wb}.  We use the central NLO LHAPDF set\cite{Butterworth:2015oua,Buckley:2014ana}, 
with
$\mu_R=\mu_F=M_{hh}$\footnote{$M_{hh}\equiv (p_{h_1}+p_{h_3})^2$.}. 
In Fig. \ref{fg:mhh}, we show the invariant $M_{hh}$ spectrum for resonant $h_1 h_3$ production compared to the SM
$h_1 h_1$ spectrum at $13~TeV$.  The complex singlet model curves are more sharply peaked than those of the SM and demonstrate a significant enhancement of the rate relative to the SM double Higgs rate
for the parameters we have chosen.  
The spectrum has only a small dependence on $\theta_2$, visible at high $M_{hh}$.  We have included a finite width for $m_2$ in the calculation:
For $m_ 2=400~GeV$ and $m_3=130~GeV$, the width is quite large, $\Gamma_2=263~GeV (\theta=0)$ and $\Gamma_2=295~GeV 
(\theta={\pi/12})$\footnote{
The parameters of the $\theta_2=0$ curve on the LHS of Fig. \ref{fg:diags} are, for example,    $\delta_2=21.6$,
     $Re(\delta_3)= -14.5$,     $Im(\delta_3)= -22.9$,     $Re(d_1)=1.15$,       $Im(d_1)=1.64$,
      $d_2= 13.3$,      $Re(d_3)= 10.5$,      $Im(d_3)= 10.2$,    $Re(e_1)= 1.18v$,
           $Im(e_1)= -2.66v$,     $Re(e_2)=-8.29v$,     $Im(e_2)= 3.67v$.  These parameters correspond to 
                  $\lambda_{211}=  -2.9v$     $\lambda_{311}= 6.77v$,       
                  $ \lambda_{321}=  -11.1v$ and     $ \lambda_{331}= 11.2v$.}. We have included the width using the Breit-Wigner approximation, although
                  typically $\Gamma_2/m_2\sim {\cal{O}}({1\over 2})$.
The shoulder due to the width is clear on the LHS of Fig. \ref{fg:diags}.
There is a smaller width for $h_2$ when $m_3$ is increased to $250~GeV$:  $\Gamma_2=129~GeV (\theta=0)$ and $\Gamma_2=137~GeV (\theta={\pi/12})$ on the RHS of Fig. \ref{fg:diags}. The widths are calculated by scaling the SM results from Ref. \cite{Dittmaier:2011ti} with the appropriate mixing angles and
adding the relevant widths $h_i\rightarrow h_j h_k$.

\begin{figure}
  \centering
{\includegraphics[width=0.48\textwidth]{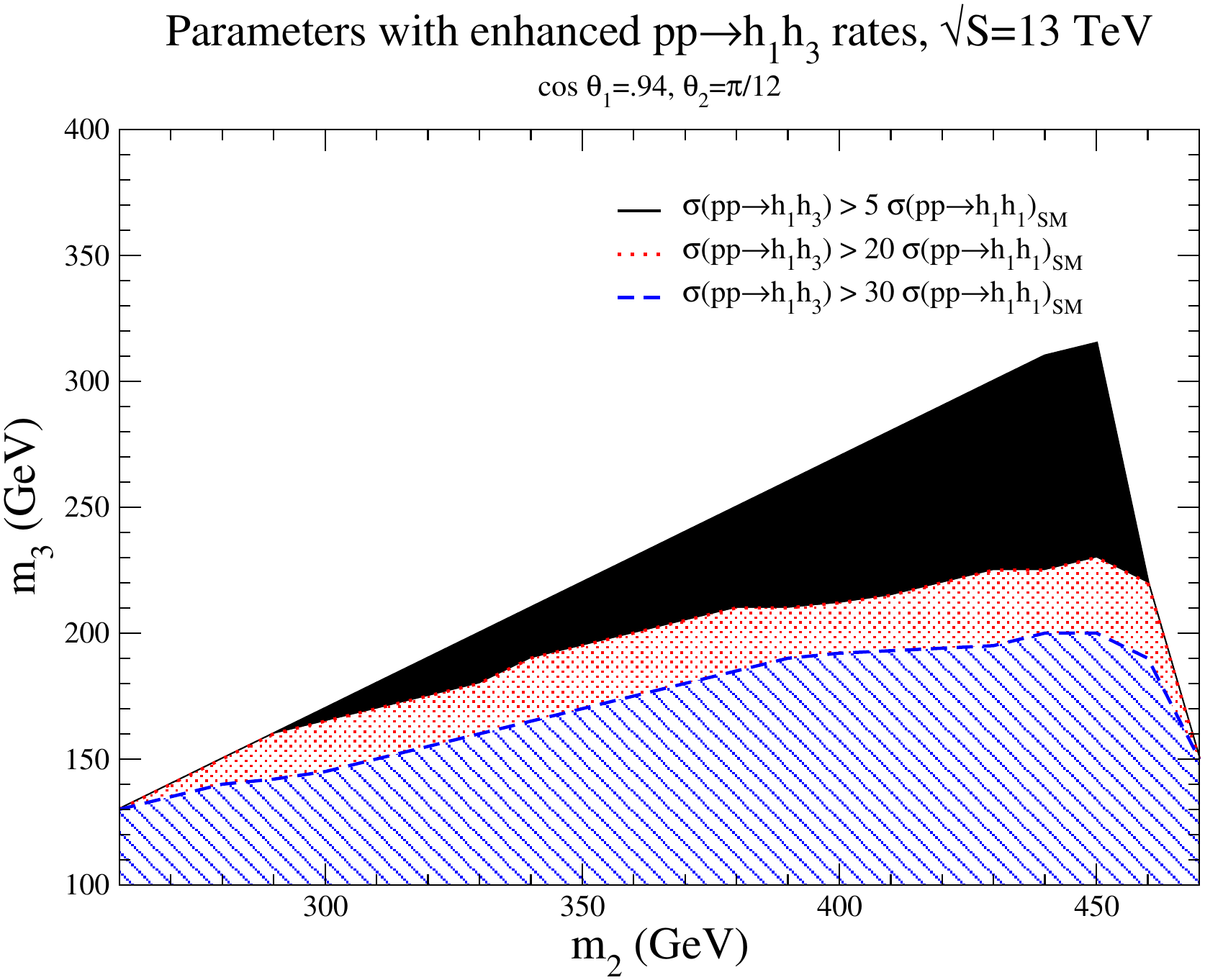}}
{\includegraphics[width=0.48\textwidth]{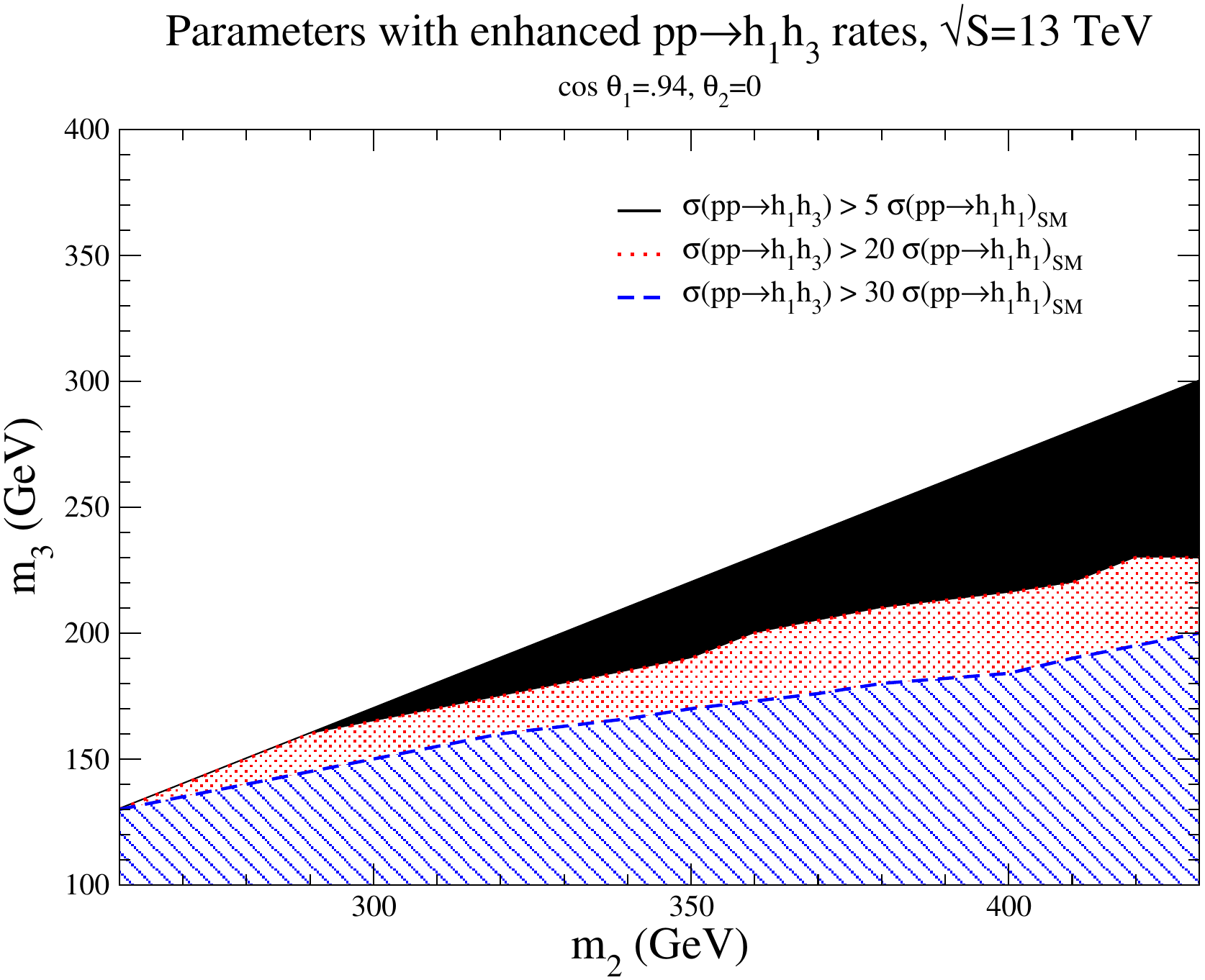}}
 \caption{Regions of parameter space allowed by limits on oblique parameters, perturbative unitarity, and the minimization of 
 the potential where the rate for $h_1h_3$ production is significantly larger than the SM $h_1h_1$ rate at $\sqrt{S}=13~TeV$.
 \label{fg:rates}
 }
 \end{figure}

\begin{figure}
  \centering
{\includegraphics[width=0.48\textwidth]{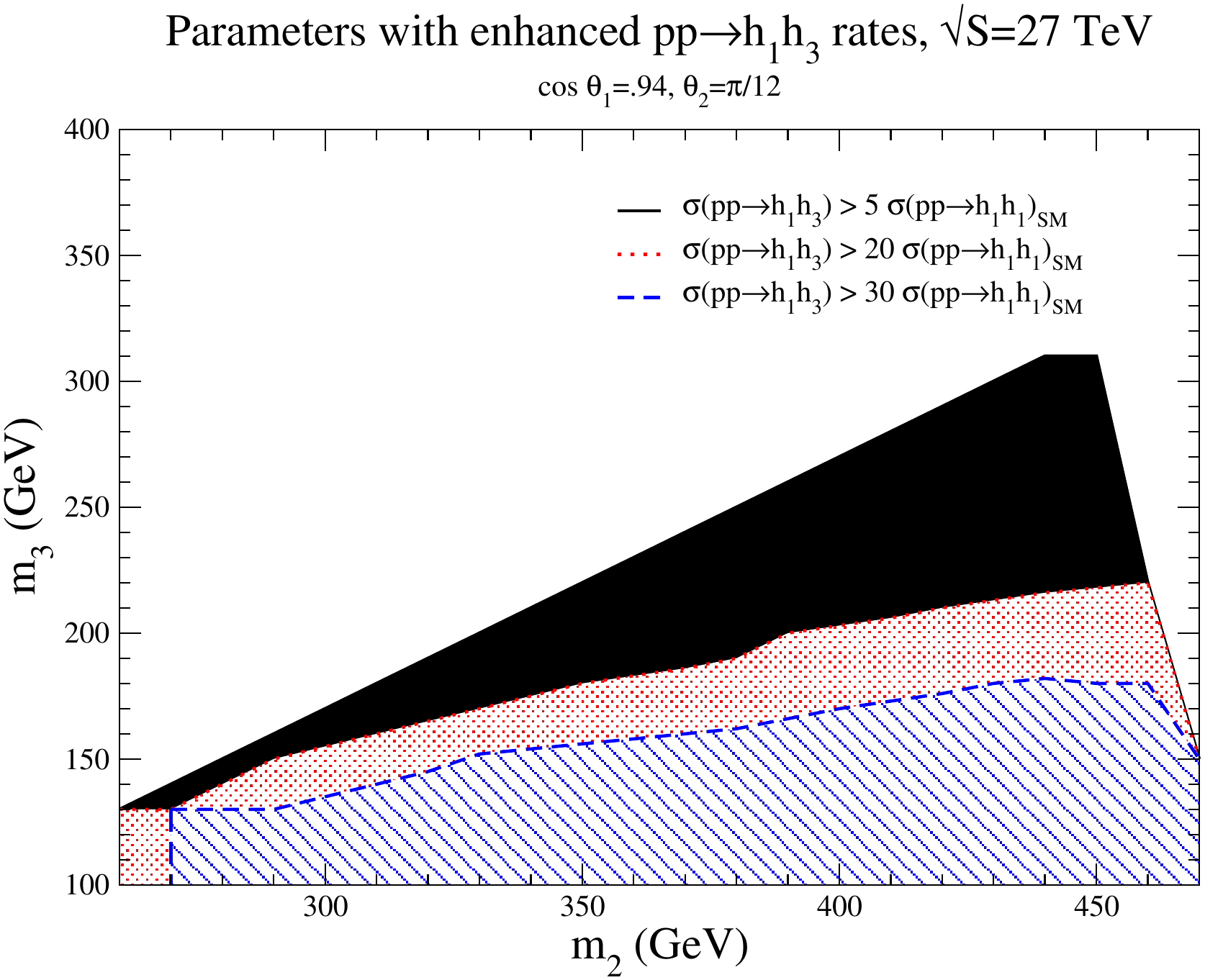}}
{\includegraphics[width=0.48\textwidth]{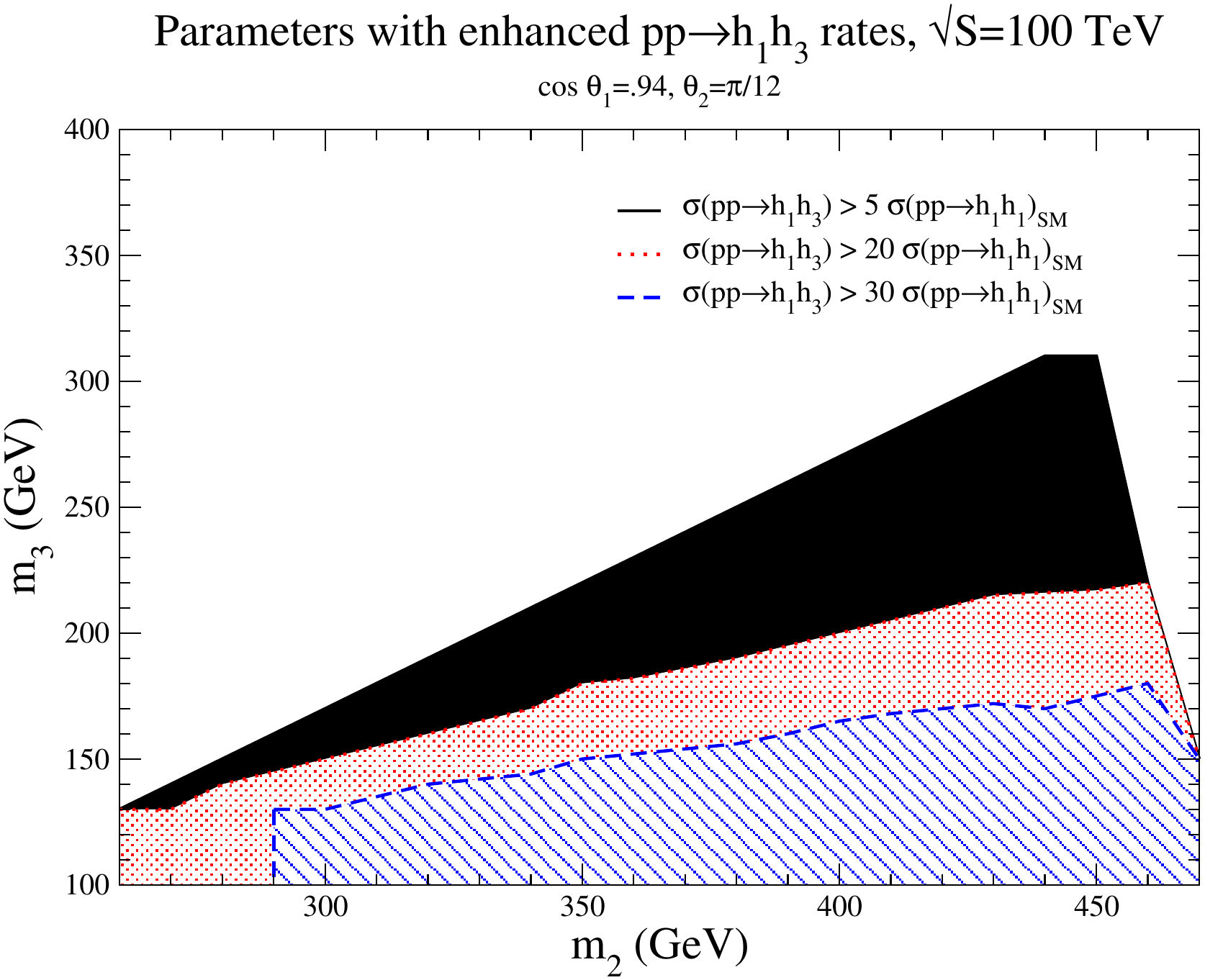}}
 \caption{Regions of parameter space allowed by limits on oblique parameters, perturbative unitarity, and the minimization of 
 the potential where the rate for $h_1h_3$ production is significantly larger than the SM $h_1h_1$ rate at $\sqrt{S}=27~TeV$
 and $100~TeV$.
 \label{fg:rates27}
 }
 \end{figure}

\begin{figure}
  \centering
{\includegraphics[width=0.48\textwidth]{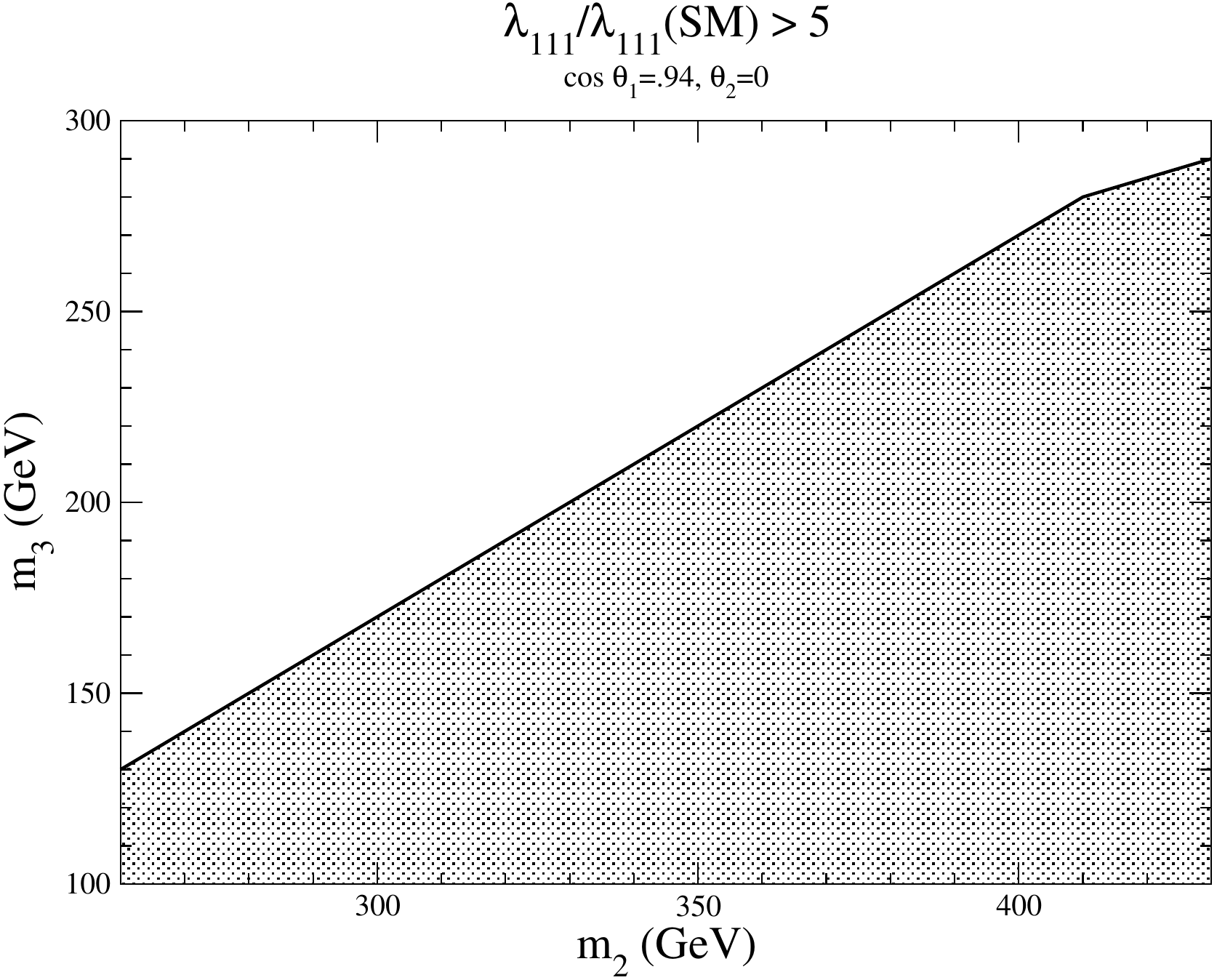}}
 \caption{Region of parameter space allowed by limits on oblique parameters, perturbative unitarity, and the minimization of 
 the potential where the $h_1h_1 h_1$ tri-linear coupling is greater than $5$ times the SM value.
  \label{fg:tri}
 }
 \end{figure}

In Figs. \ref{fg:rates} and \ref{fg:rates27}, we show mass regions where the rate for $h_1h_3$ production is significantly
enhanced relative to the SM $h_1h_1$ production.   This enhancement can be traced to the relatively large values of the
tri-linear Higgs couplings defined from Eq. \ref{eq:vdef},
\begin{equation}
{\cal V}\rightarrow {1\over 2}\lambda_{211}h_1^2 h_2+{1\over 2}\lambda_{311}
h_1^2h_3+{1\over 2}\lambda_{331} h_1 h_3^2 +\lambda_{321}h_1h_2h_3+\cdots\, ,
\end{equation}
that are allowed by the imposed restrictions.  In the SM, the $hhh$ coupling is fixed by $m_h$, whereas here, the trilinear 
couplings of the potential are
relatively unconstrained.  
 In Fig. \ref{fg:tri}, we show the region of parameter space allowed by limits on  the oblique parameters, perturbative unitarity, and the minimization of 
 the potential where the $h_1h_1 h_1$ tri-linear coupling is greater than $5$ times the SM value.    This enhancement of the tri-linear scalar coupling requires rather light values of $m_2$ and $m_3$
 as shown in Fig. \ref{fg:tri}. In roughly the same region as shaded in Fig. \ref{fg:tri}, the $h_2h_1h_1$ and  $h_3h_2h_1$  couplings are $8$ times the SM
 $h_1h_1h_1$ coupling.
 This enhancement is consistent with the results of Ref. \cite{Costa:2015llh} in the complex singlet model with
a  global $U(1)$ symmetry imposed on the potential.  
The cut-offs on the high $m_2$ ends of the plots on the LHS in   Figs. ~\ref{fg:rates} and \ref{fg:rates27} are due to the oblique parameter restrictions
in the  non-zero $\theta_2$ mixing scenario.  The same results for $\sqrt{S}=27$ and $100$ TeV are shown in Fig. \ref{fg:rates27}. 
 At all energies there is a 
significant region of phase space where the $h_1h_3$ rate is large, relative to SM double Higgs production.  

For $m_3>250~GeV$,
the dominant decay chain  from $h_1 h_3$ production will be  $h_1h_3\rightarrow h_1h_1 h_1\rightarrow (b{\overline b})
(  b{\overline b}) ( b{\overline b})$.   For $m_3 < 2 m_1$, $h_3$ will decay  through the extremely 
small couplings to SM particles and through the off-shell decay $h_3\rightarrow h_1 h_1^*\rightarrow h_1 f {\overline f}$ and will
be extremely long lived. 
In the limiting case where $\theta_2=0$, the only allowed decay for $h_3$ is the off-shell decay chain through the couplings to $h_1$.

\section{Conclusions}

We have studied an extension of the SM with a complex scalar singlet.  We considered the most general
renormalizable scalar potential and imposed no additional symmetries.  In this scenario, there are $3$ scalar
bosons, one of which, $h_3$, has very small couplings to  SM particles and will be primarily observed through di-Higgs decays,
$h_2\rightarrow h_1 h_3$.  Subject to the constraints of electroweak precision measurements, single Higgs production
rates, and perturbative unitarity, there are regions of phase space where the rate for $h_1h_3$ production is significantly
enhanced relative to the SM $h_1h_1$ rate.  Therefore, the search for pair production of Higgs bosons with different
masses is a distinctive signature of this class of model. 
\section*{Acknowledgements}
S.D.  is supported by the U.S. Department of Energy under grant
No.~DE-AC02-98CH10886 and contract DE-AC02-76SF00515.   M.S. is supported 
by  the U.S. Department of Energy, Office of Science, Office of Workforce Development for Teachers and Scientists, Office of Science Graduate Student Research (SCGSR) program. The SCGSR program is administered by the Oak Ridge Institute for Science and Education
(ORISE) for the DOE. ORISE is managed by ORAU under contract number DE-SC0014664.   We thank I.M. Lewis for discussions. 

\bibliographystyle{h-physrev}
\bibliography{hh}

\end{document}